\documentclass[apj]{emulateapj}
\usepackage{apjfonts}
\usepackage{graphicx}

\shorttitle{Mass transfer in RX\,J0806.3+1527}
\shortauthors{B. Willems and V.\ Kalogera}

\begin{document}

\title{Mass transfer and the period decrease in RX\,J0806.3+1527} 

\author{B. Willems and V. Kalogera}
\affil{Northwestern University, Department of Physics and Astronomy,
  2145 Sheridan Road, Evanston, IL 60208, USA}
\email{b-willems and vicky@northwestern.edu}

\submitted{Submitted to ApJ Letters, 2005 August 9}

\begin{abstract}
We examine the nature of RX\,J0806.3+1527 and show that it is possible
to reconcile the observed period decrease and X-ray luminosity with
the transfer of mass between two white dwarfs provided that: either
the system is (i) still in the early and short-lived ($\lesssim
100$\,yr) stages of mass transfer due to atmospheric Roche-lobe
overflow, or (ii) in a standard, long-term, quasi-stationary
mass-transfer phase that is significantly ($\sim 90$\%)
non-conservative and the conversion of accretion energy to X-rays is
quite inefficient. In either of the two cases and for a wide range of
physical parameters, we find that orbital angular momentum is lost
from the system at a rate that is a factor of a few ($\lesssim 4$)
higher than the rate associated with the emission of gravitational
waves. Although the physical origin of this extra angular momentum
loss is not clear at present, it should be taken into account
in the consideration of RX\,J0806.3+1527 as a verification Galactic
source for LISA.  
\end{abstract}

\keywords{Stars: Binaries: Close, Stars: White Dwarfs, Gravitational
  Waves} 

\section{Introduction}

In recent years, much attention has been devoted to the nature of the
X-ray source RX\,J0806.3+1527. The source was discovered by ROSAT in
1990 (Beuermann et al.\ 1999) and found to be variable with a period
of $\sim 321$\,s by Israel et al.\ (1999). The latter authors
tentatively ascribed the periodicity to the rotation of a white dwarf
in an intermediate polar type cataclysmic variable (see also Beuermann
et al.\ 1999; Burwitz \& Reinsch 2001; Norton, Haswell, \& Wynn
2004). The absence of a second period (corresponding to the orbital
motion) and the presence of He emission lines in an optical follow-up,
however, lead Israel et al.\ (2002) to reclassify the system as an
AM\,CVn type double degenerate. In this picture, the $\sim 321$\,s
periodicity reflects the orbital motion, making RX\,J0806.3+1527 the
tightest known binary to date (see also Burwitz \& Reinsch 2001;
Ramsay, Hakala, \& Cropper 2002) and an excellent verification source
for the gravitational-wave space mission LISA planned by NASA and ESA.

More recent observations revealed two possibly major problems for the
AM\,CVn model: (i) Israel et al.\ (2003), assuming a distance of
500\,pc, derived a {\em low} average X-ray luminosity of $\simeq 5
\times 10^{32}\,{\rm erg\,s^{-1}}$, and (ii) Hakala et al.\ (2003) and
Strohmayer (2003) found the observed period to be {\em decreasing}
with time at a rate of $\simeq 10^{-11}\,{\rm s\, s^{-1}}$ (see also
Israel et al.\ 2004; Strohmayer 2005). The X-ray luminosity poses a
problem because it is significantly below the luminosity expected from
mass accretion in compact double white dwarfs, while the period
decrease poses a problem because mass transfer in AM\,CVns is expected
to expand the orbit rather than shrink it. A third possible model was
therefore proposed to be the unipolar inductor model developed by Wu
et al.\ (2002). In this model the X-ray flux is generated by electric
currents between two close but detached white dwarfs. However,
shortcomings of this model have recently been pointed out by Barros et
al.\ (2005) and Marsh \& Nelemans (2005). The question on the nature
of RX\,J0806.3+1527 therefore still remains open.

In this {\em Letter}, we revisit the AM\,CVn model for
RX\,J0806.3+1527 and investigate under which conditions the assumption
of mass transfer (MT) between two white dwarfs can comfortably explain
all the currently available observational constraints. Our motivation
stems from both trying to understand the origin of the X-ray emission
and the measured period decrease and to examine the physical
properties of this binary and its role as a verification source for
LISA.

\section{The mass and radius of the donor star}
\label{mr}

We assume RX\,J0806.3+1527 to be an AM\,CVn type double degenerate
with an orbital period $P_{\rm
orb}=5.4$\,min ($\sim 321$\,s).  We denote the
masses of the two white dwarfs (WD) by $M_1$ and $M_2$, and assume
$M_1 > M_2$ so that the primary corresponds to the mass accretor and
the secondary to the mass donor. Since the evolutionary channels leading to
the formation of AM\,CVns involve multiple MT phases (e.g., Han 1998,
Nelemans et al.\ 2001a), the orbit can be safely assumed circular.

\begin{figure}
\resizebox{\hsize}{!}{\includegraphics{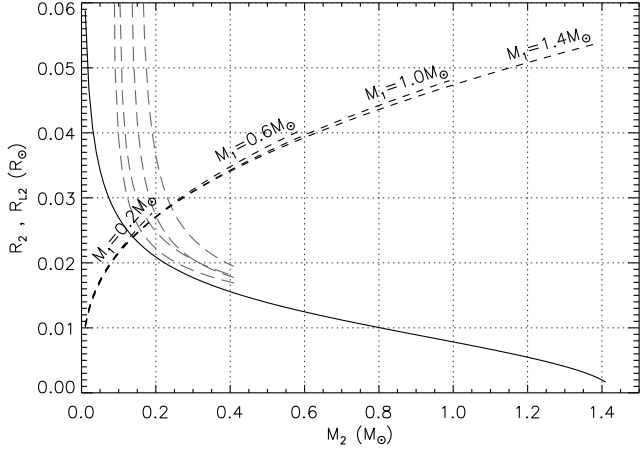}} 
\caption{Mass-radius relations for zero-temperature (solid line) and
  finite temperature (long-dashed lines) WD models. From left to
  right, the finite temperature models correspond to $(M_{\rm
  H}/M_\odot, T_2/K) = (10^{-6},5 \times 10^3), (3 \times 10^{-4},5
  \times 10^3)$, $(10^{-6},10^4)$, $(3 \times 10^{-4},10^4)$. The
  donor Roche-lobe radii for $P_{\rm orb}=5.4$\,min, $M_1=0.2, 0.6,
  1.0, 1.4\,M_\odot$, and $M_2 < M_1$ are also shown (short-dashed
  lines).}
\label{rlo}
\end{figure}

The assumptions that the observed periodicity of 5.4\,min corresponds
to the orbital period and that the secondary fills its critical Roche
lobe can be used to constrain the mass $M_2$ and radius $R_2$ of the
donor. For this purpose, we first determine the radius of the
secondary as a function of its mass by means of the mass-radius
relation for zero-temperature WDs (Nauenberg 1972):
 \begin{equation}
 R_2 = 0.01125\, R_\odot \left[ \left( {M_2 \over M_{\rm ch}}
   \right)^{-2/3} - \left( {M_2 \over M_{\rm ch}} \right)^{2/3}
   \right]^{1/2},  \label{R2}
 \end{equation}
 where $M_{\rm ch}$ is the Chandrasekhar mass (see also Han \& Webbink
 1999). The resulting radii are shown in Fig.~\ref{rlo} (solid
 line). Next, we consider the mass-radius relations for finite
 temperature helium WDs with masses from 0.1 to 0.4\,$M_\odot$ derived
 by Hansen \& Phinney (1998). Radii for effective temperatures $T_2 =
 5 \times 10^3$ and $10^4$\,K, and hydrogen envelopes of $M_{\rm H} =
 10^{-6}$ and $3 \times 10^{-4}\, M_\odot$ are shown in Fig.~\ref{rlo}
 (long-dashed lines). They are systematically larger than radii for
 zero-temperature WDs, with the largest differences occurring at the
 lowest masses.

Under the assumption that the secondary is not significantly out of
thermal equilibrium, its radius is approximately equal to the volume-equivalent
radius $R_{\rm L2}$ of its critical Roche lobe (see Eggleton 1983).
The variations of the latter as a function of $M_2$ 
are shown in Fig.~\ref{rlo} (short-dashed lines) for $P_{\rm
orb}=5.4$\,min and $M_1 = 0.2, 0.6, 1.0, 1.4\,M_\odot$.
It follows that, for zero-temperature WD models, $R_2 = R_{\rm L2}$ for $M_2 = 0.13\,M_\odot$ and $R_2 =
0.024\,R_\odot$ (cf. Israel et al.\ 2002). Finite-temperature
mass-radius models yield larger donor masses and radii. For $T_2
= 10^4$\,K and $M_{\rm H}=3 \times 10^{-6}\, M_\odot$, e.g., $M_2 =
0.24\,M_\odot$ and $R_2 = 0.029\,R_\odot$.

\section{The mass-transfer rate and X-ray luminosity}
\label{MT}

In AM\,CVn binaries, the observed X-ray luminosity is generated by
mass accretion onto the primary. Ramsay et al.\ (2005) have shown that
a large fraction of the accretion luminosity in AM\,CVns is radiated
in the UV rather than in X-rays. In particular, they estimated the
ratio of the X-ray luminosity to the UV luminosity to be $\sim 1/1000$
for AM\,CVn and $\sim 1/3$ for GP\,Com. In order to accommodate this,
we set $L_{\rm X} = \alpha\, L_{\rm acc}$, where $L_{\rm X}$ and
$L_{\rm acc}$ are the X-ray and total accretion luminosity,
respectively, and $0 \le \alpha \le 1$. We {\em conservatively} set
$\alpha=0.1$ and we note that $\alpha$ can very well be lower by 1-2
orders of magnitude. The linear dependence of $L_{\rm X}$ on $\alpha$
allows for an easy rescaling of our results for different values of
$\alpha$.

The accretion luminosity due to MT is given by
\begin{equation}
L_{\rm acc} = -\beta\ \dot{M}_2 \left( \Phi_{\rm L1} 
   - \Phi_{\rm R1} \right),  \label{Lacc}
\end{equation}
where $\beta$ is the fraction of the transferred mass accreted by
the primary, and $\Phi_{\rm L1}$ and $\Phi_{\rm R1}$ are the Roche
potential values at the inner Lagrangian point and at the surface of
the accretor, respectively (see Han \& Webbink 1999 for details, but
note the reverse definition of $M_1$ and $M_2$). To determine
$\Phi_{\rm R1}$, we model the accretor as a zero-temperature white
dwarf, noting that finite temperatures lead to an increase in the
radius and thus a decrease in the absolute value of $\Phi_{\rm R1}$
and $L_{\rm acc}$.

\subsection{Quasi-Stationary Mass Transfer} 

In order to estimate the accretion rate and X-ray luminosity, we need
to determine $\dot{M}_2$ from the donor. We assume that MT turns on
instantaneously when $R_2=R_{L2}$ and that $\dot{M}_2$ is given by the
quasi-stationary rate
\begin{equation}
\dot{M}_2 = -{M_2 \over {\zeta_{\rm s2} - \zeta_{\rm L2}}}
   \left( {\dot{R}_2 \over R_2} - 2\, 
   {\dot{J}_{\rm orb} \over J_{\rm orb}} \right). \label{qs}
\end{equation}
Here $J_{\rm orb}$ is the orbital angular momentum, $\dot{R_2}$ and
$\dot{J}_{\rm orb}$ correspond to time derivatives of $R_2$ and
$J_{\rm orb}$ at constant mass, $\zeta_{\rm s2}=(\ln R_2/\ln M_2)_s$
is the donor's adiabatic radius-mass exponent, and $\zeta_{\rm
L2}=(\ln R_{\rm L2}/\ln M_2)$ (e.g. Rappaport, Joss, \& Webbink
1982; Ritter 1988; Han \& Webbink 1999). Expressions for $\zeta_{\rm
s2}$ and $\zeta_{\rm L2}$ appropriate for WDs are given by Han \&
Webbink (1999)\footnote{The $\zeta_{\rm s2}$ expression derived by Han
\& Webbink (1999) is valid only for zero-temperature WDs [$M_2$-$R_2$
relation as in Eq.~(\ref{R2})]. In our finite-temperature WD
calculations we therefore use a generalization of $\zeta_{\rm s2}$
based on the $M_2$-$R_2$ relations of Hansen \& Phinney (1998). The
adopted expression for $\zeta_{\rm L2}$ is derived under the
assumption that any mass lost from the system carries away the
specific orbital angular momentum of the accretor.}.  For our purpose,
we neglect changes in the stellar radius related to the cooling of the
white dwarf at constant mass, and assume the orbital evolution in the
absence of MT to be driven entirely by orbital angular momentum losses
due to gravitational radiation (GR). In Eq.~(\ref{qs}), we then have
$\dot{R}_2 = 0$ and $\dot{J}_{\rm orb} = \dot{J}_{\rm GR}$.

\begin{figure}
\resizebox{\hsize}{!}{\includegraphics{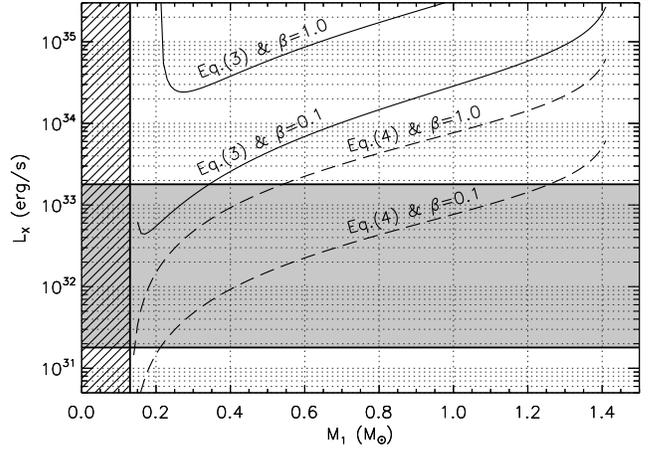}}
\caption{Variations of $L_X$ as a function of $M_1$, for
  zero-temperature WDs, $M_2=0.13\,M_\odot$, $\alpha=0.1$, and
  $\beta=0.1$ or 1. Solid lines correspond to the quasi-stationary MT
  rate with $\dot{J}_{\rm orb} = \dot{J}_{\rm GR}$ [Eq.~(\ref{qs})];
  dashed lines to the turn-on rate [Eq.~(\ref{m2dot})]. The grey
  horizontal band indicates the observationally inferred $L_X$ for
  distances of 100--1000\,pc. Accretors with $M_1 < 0.13\,M_\odot$
  (hatched area) are excluded because of the $M_1 > M_2$ assumption.}
\label{lx}
\end{figure}

For zero-temperature WDs (i.e. $M_2=0.13\,M_\odot$), we calculate
$\dot{M}_2 \sim 10^{-7}$--$10^{-5}\,M_\odot\, {\rm yr^{-1}}$ for
conservative MT ($\beta=1$), and $10^{-7}$--$10^{-6}\,M_\odot\, {\rm
yr^{-1}}$ for non-conservative MT ($\beta=0.1$). The associated X-ray luminosities 
are shown in Fig.~\ref{lx} as functions of $M_1$ (solid lines).  For
$\beta=1$, $L_{\rm X}$ is well outside the grey horizontal band
indicating the observational constraints on $L_{\rm X}$ for distances
between 100 and 1000\,pc (Israel et al.\ 2002, Marsh \& Nelemans
2005). For $\beta=0.1$, the calculated $L_{\rm X}$ is consistent with
the observed one if $0.13 \la M_1/M_\odot \la 0.35$, a
very narrow range of accretor masses\footnote{Equation~(\ref{Lacc}) is
often approximated by $L_{\rm acc} = -\beta G M_1
\dot{M}_2/R_1$. However, as noted by Han \& Webbink (1999), this is
inappropriate for compact double WDs in which the donor is located
deep within the potential well of the accretor. For the system
parameters considered here, the use of the approximation yields $L_X$
values up to a factor of $\sim 5$ larger than those shown in
Fig.~\ref{lx}.}. For a given value of $\beta$, finite temperature WDs
give systematically higher MT rates and thus higher $L_X$ values.

We conclude that the quasi-stationary MT rate [Eq.~(\ref{qs})] yields
$L_X$ values that are only marginally compatible with the observed
range, {\em unless} $\alpha \simeq 1/1000$ (as for AM\,CVn) {\em and} $\beta \simeq
0.1$.  However, it should be noted that the quasi-stationary rate is
only representative of the long-term average MT rate. Marsh \&
Nelemans (2005) estimated that deviations from the quasi-stationary
rate can take place on time scales of $\sim 100$\,yrs, suggesting that
below-average MT rates could be sustained for time spans sufficiently
longer than the 15 years since the discovery of the source. The
mechanism responsible for driving the MT rate away from its
equilibrium value remains unclear, but, as noted by Marsh \& Nelemans
(2005), a comforting list of candidates exist, although their
applicability to double WDs has yet to be investigated.

\subsection{Atmospheric Roche-lobe Overflow}

Alternatively, the MT rate in RX\,J0806.3+1527 could also be below its
long-term average if the system started MT only recently, and
therefore is in a stage of atmospheric Roche-lobe
overflow (RLO) and has yet to evolve
towards the quasi-stationary state. A modified rate applicable to
turn-on MT was derived by Ritter (1988) who pointed out the role of
the atmospheric properties of the donor star during the turn-on phases
(see also D'Antona, Mazzitelli, \& Ritter 1989; Kolb \& Ritter
1990). We therefore reconsider the $L_{\rm X}$ derivation under the
assumption that MT started only recently.

Following Ritter (1988), we determine 
\begin{equation}
\dot{M}_2 = - {{2\,\pi} \over \sqrt{e}} \left( {{{\cal R}\, T_2} 
   \over \mu_2} \right)^{3/2} {R_{\rm L2}^3 \over {G\, M_2}}\, 
   \rho_2\, F_2(q)\, \exp \left( {{R_2 - R_{\rm L2}} 
   \over H_{P2}} \right),  \label{m2dot}
\end{equation}
with an associated exponential turn-on time scale
\begin{equation}
\tau_{\dot{M}_2} = - {1 \over 2}\, {H_{P2} \over R_2}\, 
   {J_{\rm orb} \over \dot{J}_{\rm orb}}.  \label{tau}
\end{equation}
Here $G$ is the Newtonian constant of gravitation, ${\cal R}$ the
universal gas constant, $T_2$ the effective temperature, $\rho_2$ the
photospheric mass density, $\mu_2$ the photospheric mean molecular
weight, and $H_{P2}$ the photospheric pressure scale
height. Furthermore, $F_2(q)$ is a function of the mass ratio
$q=M_1/M_2$ which, for $0.5 \la q \la 10$, can be approximated as
$F_2(q) = 1.23 + 0.5\, \log q$.

From Eq.~(\ref{m2dot}), it is clear that, although convenient for the
derivation of mass-radius relations, the zero-temperature
approximation is not suitable when considering the turn-on MT rate in
semi-detached double WDs.  Instead we determine $\dot{M}_2$ for a WD
donor with a finite but low effective temperature $T_2=2500$\,K. We
set the mean molecular weight $\mu_2=4$, as appropriate for a
non-ionized helium-rich gas, and adopt the equation of state of an
ideal gas to determine the mass density $\rho_2$ from the pressure
$P_2$ and temperature $T_2$. From the temperature-pressure
stratifications for cool white dwarfs derived by Saumon \& Jacobson
(1995) and Rohrmann (2001), it then follows that $P_2 \simeq 10^9\,
{\rm dyn\, cm^{-2}}$ and $\rho_2 \simeq 2 \times 10^{-2}\, {\rm
g\,cm^{-3}}$. The corresponding pressure scale height for
$M_2=0.13\,M_\odot$ and $R_2=0.024\,R_\odot$ is $H_{P2} \simeq 5
\times 10^{-6}\,R_2$. Setting $R_2 = R_{\rm L2}$ then yields MT rates
$\dot{M}_2 \simeq 5 \times 10^{-9}\,M_\odot\,{\rm yr^{-1}}$ and
exponential turn-on time scales $\tau_{\dot{M}_2} \simeq
2$--12\,yrs. The corresponding X-ray luminosities are shown in
Fig.~\ref{lx} (dashed lines). They are considerably lower than those
obtained using the quasi-stationary MT rate and are consistent with
the observed $L_X$ for a wide range of $M_1$ values, especially for
$\beta\simeq 0.1$. We note however that if $\alpha\sim1/1000$, 
the $\beta\sim1$ case is more favorable.

The derivation for WDs of $5 \times 10^3$\,K and $10^4$\,K yields
$\dot{M}_2 \simeq 2 \times 10^{-10}$ and $\simeq
10^{-12}\,M_\odot\,{\rm yr^{-1}}$, and turn-on time scales of $\simeq
5$--15 and $\simeq 5$--20\,yrs, respectively. The associated $L_X$ are
therefore even lower than those shown in Fig.~\ref{lx}.

\section{The orbital angular momentum loss rate}

The AM\,CVn model for RX\,J0806.3+1527 implies that the orbital period
should be increasing in time, unless additional orbital angular
momentum loss mechanisms besides GR are operating in the system; or if
MT started so recently that the accompanying orbital expansion has yet
to overtake the orbital contraction driven by GR. Exploration of the
latter possibility requires detailed MT calculations using realistic
WD models, which is beyond the scope of this investigation. Instead,
we focus on the additional orbital angular momentum losses required to
reconcile the predicted orbital evolution with the measured period
decrease of $3.6\times 10^{-11}\,{\rm s\, s^{-1}}$ (Strohmayer
2005). We tentatively ascribe the additional orbital angular momentum
losses to spin-orbital coupling through tides and/or magnetic fields,
although our analysis is generally valid for other mechanisms as well.

\begin{figure}
\resizebox{\hsize}{!}{\includegraphics{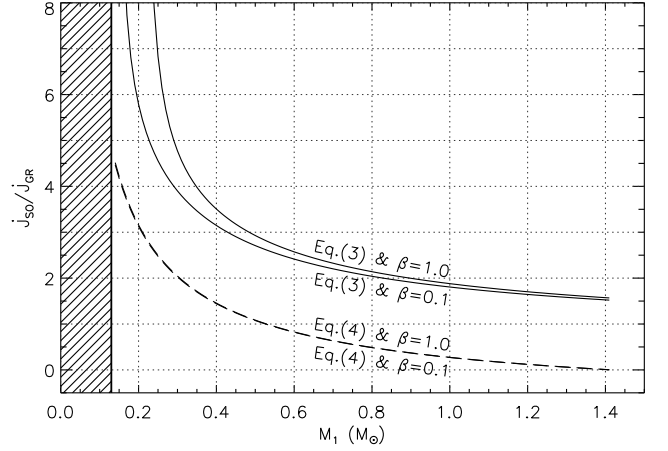}}
\caption{The orbital angular momentum loss rate due to spin-orbit
  coupling required for the AM\,CVn model to be compatible with the
  observed period decrease in RX\,J0806.3+1527. Solid lines
  correspond to the quasi-stationary MT rate for zero-temperature WDs
  [Eq.~(\ref{qs})]; dashed lines to the turn-on rate for $T_2=2500$\,K
  WDs [Eq.~(\ref{m2dot})]. In the latter case, the curves associated
  with $\beta=0.1$ and 1 are almost indistinguishable because of the
  small contribution of the $\dot{M}_2$ terms in
  Eqs.~(\ref{jdot2})--(\ref{jdotMT}. Both the
  zero-temperature and $T_2=2500$\,K calculations are based on a donor
  mass of $0.13\,M_\odot$. The hatched region is excluded for the same
  reason as in Fig.~\ref{lx}.}
\label{jorbdot}
\end{figure}

From the definition of $J_{\rm orb}$ and Kepler's third law, it
follows that 
\begin{equation}
{\dot{J}_{\rm orb} \over J_{\rm orb}} = {1 \over 3}\, 
   {\dot{P}_{\rm orb} \over P_{\rm orb}} + \left[ {1 \over M_2} 
   - {\beta \over M_1} - {{1 - \beta} \over 
   {3 \left( M_1 + M_2 \right)}} \right] \dot{M}_2.  \label{jdot2}
\end{equation}
Writing $\dot{J}_{\rm orb}$ as the sum of the contributions from
gravitational radiation (GR), spin-orbit coupling (SO), and systemic
mass loss due to non-conservative mass transfer (MT), on the other
hand, yields
\begin{equation}
\dot{J}_{\rm orb} = \dot{J}_{\rm GR} + \dot{J}_{\rm SO} 
   + \dot{J}_{\rm MT}.  \label{jdot}
\end{equation}
The first two terms in Eq.~(\ref{jdot}) operate in both detached and
semi-detached double WDs. The third term becomes active only after the
onset of MT. We model this term assuming that any systemic mass loss
carries away the specific orbital angular momentum of the accretor,
i.e.
\begin{equation}
\dot{J}_{\rm MT} = \left( 1 - \beta \right)\, {M_2 \over M_1}\, 
   {J_{\rm orb} \over {M_1+M_2}}\, \dot{M}_2.  \label{jdotMT}
\end{equation}

Substitution of Eqs.~(\ref{jdot2}) and~(\ref{jdotMT}) into
Eq.~(\ref{jdot}), and use of the well known expression for
$\dot{J}_{\rm GR}$ (e.g Landau \& Lifshitz 1962) and the observed
period and rate of period decrease yields the $\dot{J}_{\rm SO}$
values required for consistency with observations. For
zero-temperature WDs ($M_2=0.13\,M_\odot$), $\dot{J}_{\rm
SO}$ is shown in Fig.~\ref{jorbdot} in units of $\dot{J}_{\rm GR}$.
In the case of the quasi-stationary MT rate [Eq.~(\ref{qs})],
$\dot{J}_{\rm SO} \simeq 1.5\, \dot{J}_{\rm GR}$ for $M_1 \simeq
1.4\,M_\odot$, and $\dot{J}_{\rm SO} \ga 3.5\, \dot{J}_{\rm GR}$ for
$M_1 \la 0.35\,M_\odot$ (the $M_1$ range where the calculated $L_{\rm
X}$ for $\alpha=0.1$ and $\beta=0.1$ is consistent with the observed
one). For $5 \times 10^3$\,K and $10^4$\,K WDs, $\dot{J}_{\rm SO}$ is
even larger. On the other hand, in the case of the turn-on MT rate
[Eq.~(\ref{m2dot})], $\dot{J}_{\rm SO} \la 4.5\, \dot{J}_{\rm GR}$
over the entire $M_1$ range, and $\dot{J}_{\rm SO}$ is negligible
($\simeq 10^{-3}\, \dot{J}_{\rm GR}$) for $M_1 \simeq
1.4\,M_\odot$. This also applies to finite WD temperatures because of
the low MT rates during the onset of RLO.

\section{Discussion and conclusions}

We have revisited the AM\,CVn model for RX\,J0806.3+1527 and
confronted its theoretical predictions with the observational
constraints. For an orbital period of 5.4\,min and a donor temperature
of less than $10^4$\,K, RLO from the secondary occurs for $M_2 \simeq
0.13$--$0.24\,M_\odot$ and $R_2 \simeq 0.024$--$0.029\,R_\odot$. Under
the assumptions that 10\% of the transferred mass is accreted by the
companion, and that 10\% of accretion luminosity is radiated in
X-rays, the X-ray luminosity due to accretion is found to be
comfortably consistent with the observed X-ray luminosity within the
distance uncertainties ($d \la 1$\,kpc).

We have examined both long-term quasi-stationary and short-term
atmospheric MT with exponential growth. We found that the
predicted $L_X$ and $\dot{J}_{\rm orb}$ are more comfortably
reconciled with the observations in the latter of the two cases. The
lower MT rates for atmospheric RLO yield lower accretion
luminosities, which significantly alleviates the constraints on the
fraction of the mass accreted by the companion and the fraction of the
accretion luminosity emitted in X-rays. The predicted orbital
evolution is compatible with the measured period decrease, if orbital
angular momentum losses other than just GR (e.g.\ due to spin-orbit
coupling) generally play a non-negligible role. We find that, in the
case of quasi-stationary MT, an extra loss rate of at least
a factor of $\simeq1.5$ times the GR rate is required. On the other
hand, in the case of atmospheric MT, this factor is smaller
than $4.5$ over the entire $M_1$ range and vanishes altogether for
$M_1 \ga 1.2$\,M$_\odot$. The drawback of the atmospheric RLO
hypothesis is that MT must have started only very recently
(within $\simeq5\times\tau_{\dot{M}_2}$, i.e., at most $\sim100$\,yr).

The standard hypothesis of quasi-stationary MT, however,
can also become consistent with observations if MT is
significantly non-conservative and the fraction of the accretion
luminosity emitted in X-rays (instead of UV) is as low ($\simeq
1/1000$) as estimated by Ramsay et al.\ (2005) for AM\,CVn. The
drawback of this hypothesis is that the required additional orbital
angular momentum losses are about $1.5-4$ times the GR loss rate (for
$M_1 \ga 0.3$\,M$_\odot$). Without a detailed study of spin-orbit
coupling through tides and/or magnetic fields it is not obvious
whether such high loss rates are sustainable.

Besides spin-orbit coupling, the rate of orbital angular momentum
loss can also be increased if matter leaving the system is assumed to
carry away the specific orbital angular momentum of the donor rather
than that of the accretor. In addition, if RX\,J0806.3+1527 is a
direct impact accretor, the absence of an accretion disk provides yet
another orbital angular momentum sink since no angular momentum from
the accreted material can be directly fed back into the orbit 
(Marsh, Nelemans, \& Steeghs 2004). We also note that in the case of quasi-stationary MT, these extra orbital angular momentum losses tend to increase $\dot{M}_2$ and $L_X$ [see Eqs.~(\ref{Lacc})--(\ref{qs})]. For $M_1 \ga 0.3$\,M$_\odot$ the increase is smaller though than the orders of magnitude uncertainty in the fraction of the accretion luminosity emitted in X-rays.

We conclude that, in view of the uncertainties in the WD temperatures,
the MT mechanism, the fraction of the accretion luminosity
radiated in X-rays, and the acting orbital angular momentum loss
mechanisms, the observed X-ray luminosity and rate of period decrease
cannot be used to exclude that RX\,J0806.3+1527 is an AM\,CVn type
double degenerate. The standard association of accretion and X-ray
emission does not at all appear inconsistent with the current
observations and hypotheses about the origin of the X-rays being
unrelated to accretion are not necessary.

The present study reveals the possibility of additional angular
momentum loss rates due to spin-orbit coupling in double WDs and
clearly indicates that the use of these interacting binaries in LISA
detection and data analysis may not be as straightforward as
previously thought.

\acknowledgments 
We thank Gijs Nelemans for discussions and sharing the results from
Marsh \& Nelemans (2005) prior to publication; Chris Deloye, Ulrich
Kolb, Sterl Phinney and Ron Taam for useful discussions on (accreting)
white dwarfs; and the Aspen Center for Physics for hospitality and
support provided during the LISA Data: Analysis, Sources, and Science
Workshop where this research was initiated. BW also acknowledges
support from the NASA Award NNG05G106G to travel to the Aspen Center
for Physics. This work is partially supported by a NSF Gravitational
Physics grant, a David and Lucile Packard Foundation Fellowship in
Science and Engineering grant, and NASA ATP grant NAG5-13236 to
VK. The authors furthermore made extensive use of NASA's Astrophysics
Data System Bibliographic Services.

\end{document}